\documentclass{article}
\usepackage[utf8]{inputenc}
\usepackage[letterpaper, portrait, margin=1in]{geometry}
\usepackage{graphicx}
\usepackage{neuralnetwork}
\usepackage{siunitx}
\usepackage{authblk}
\DeclareUnicodeCharacter{FF0C}{,}
\title{Sector Rotation by Factor Model and Fundamental Analysis}

\author[1]{Runjia Yang}
\affil[1]{University of California, Davis}
\author[2]{Beining Shi}
\affil[2]{University of California, Davis}

\date{Sept 2023}

\begin{document}

\maketitle

\section*{Abstract}
This study presents an analytical approach to sector rotation, leveraging both factor models and fundamental metrics. We initiate with a systematic classification of sectors, followed by an empirical investigation into their returns. Through factor analysis, the paper underscores the significance of momentum and short-term reversion in dictating sectoral shifts. A subsequent in-depth fundamental analysis evaluates metrics such as PE, PB, EV-to-EBITDA, Dividend Yield, among others. Our primary contribution lies in developing a predictive framework based on these fundamental indicators. The constructed models, post rigorous training, exhibit noteworthy predictive capabilities. The findings furnish a nuanced understanding of sector rotation strategies, with implications for asset management and portfolio construction in the financial domain.

\vspace{4mm}
\noindent
{\bf Keywords:} US Industrial Sectors, Factor Analysis, Fundamental Analysis, Trading Strategy.

\section{Introduction}
Sector is composed by a basket of stocks that representing companies in certain business class, which has unique features according to the business. Under certain conditions, such as economic cycles, sectors may behave accordingly due to the different characteristics of businesses. In this report, we are exploring how to capture returns by finding the hidden features behind different sectors and determining the leading sectors in some particular market conditions or social environments. Generally, this report covers a brief exploration of market and fundamental factors, explaining the meaning of each factors and how they are related to some sectors.Then we applied a neural network model to do a classification and prediction using the fundamental factors as inputs. At the end of the report, we also covers how sectors behaved under global events.

\section{Sector Classification and Return Analysis}
There are many different ways to divide sectors. For the purpose of common acceptance and convenience for future data acquirement, we used the MSCI Global Industry Classification Standard, which includes 11 level one sectors, 24 level two industry groups, 69 level three industries, and 158 sub-industries. We use the 11 level one sectors as our main target. They are Energy, Materials, Industrials, Consumer Discretionary, Consumer Staples, Health Care, Financials, Information Technology, Communication Services, Utilities, and Real Estate. In order to track the performance of each sector, we use the S$\&$P500 GICS Indices which are constructed exactly as the MSCI classification.Before working on any strategies further, we need to determine if there are actually possible profits. In our case, we need to check how big the differences between sectors' returns are. For each observation time period,
{\textbf{Define:}}  Return Difference =($ \sum_{\text{top 3}}$ Sector Return - $\sum_{\text{bottom 3}}$  Sector Return )/3
Based on a monthly frequency, we calculate the return difference and get the following plot.
\begin{center}
\includegraphics[scale=0.35]{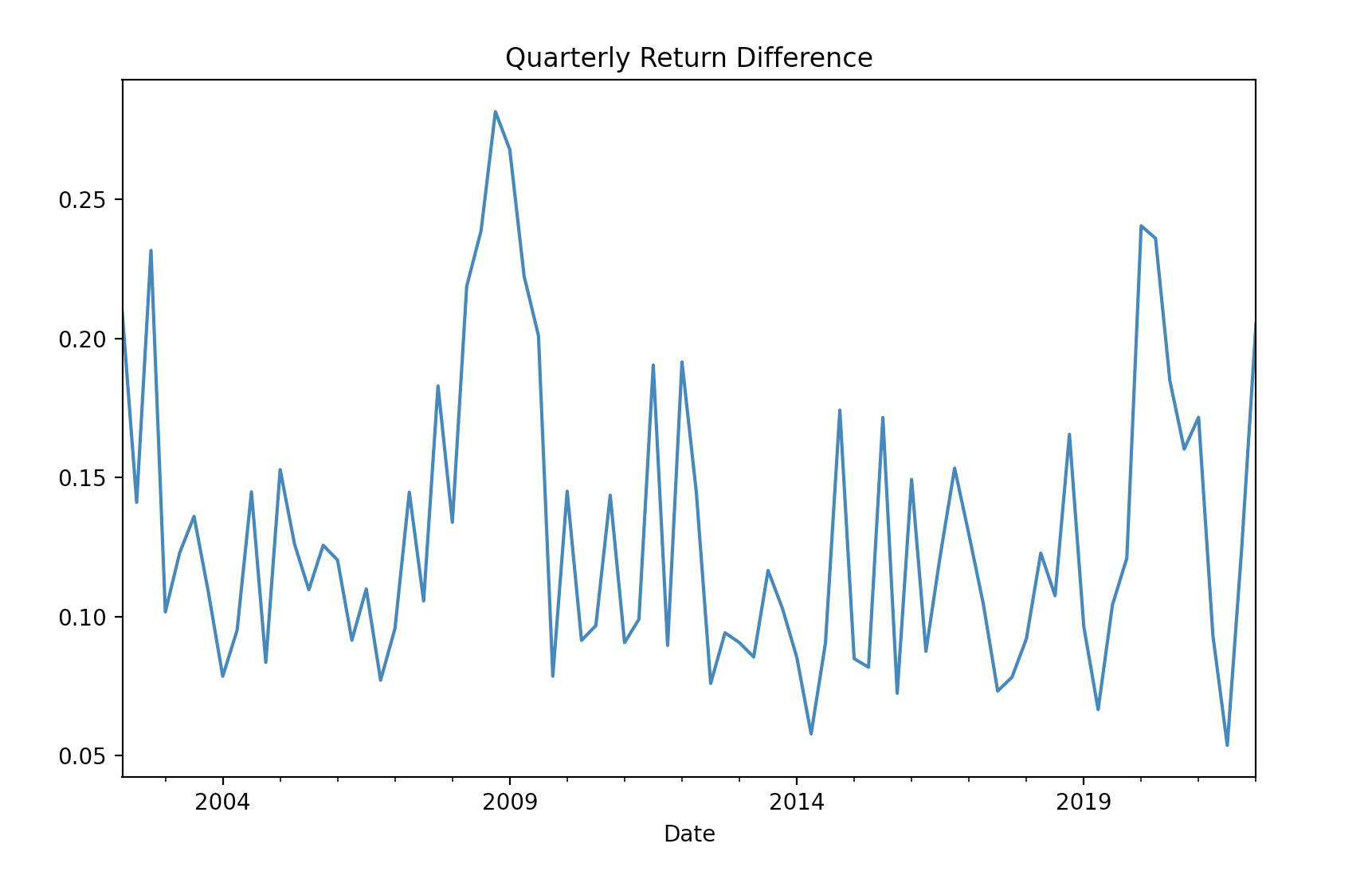}
\end{center}
Also, by calculation, the quarterly return difference has a mean of 0.1306, median of 0.1185, standard deviation of 0.0523. We can reach to a result that half of the quarterly return difference is more than 11.85$\%$. It is easy to see that there does exist potential investment opportunity by capturing the return differences between sectors. 

\section{Factor Analysis}

\subsection{Momentum Factor Exploration}
Momentum premium was first recognized by UCLA scholars Narasimhan Jegadeesh and Sheridan Titman in 1993. The momentum premium is established on the observation that assets that have performed well in the past have the trend to persist good performance in the future. Though the momentum effect is considered to be a market anomaly, it has been recognized widely among many asset classes. We will explore the momentum effect based on the sector indices introduced above.

\subsubsection{Factor Construction}
First of all, we need to construct the momentum factor. Typically, the momentum factor is constructed by the past 6 or 12 months cumulative return and excludes the most recent month's return, considering that there are also short-term reversion effects based on the mean-reversion effects. However, without a clear idea of how the sector indices carry the momentum effect, we need to explore through time intervals to find the best possible momentum factor. Then we constructed 12 different momentum factors using the past 1 to 12 month's return and excluding the most recent 0.1 portion trading days of each time period to avoid short-term reversion. For each of the factors with the period of n months
\begin{equation*}
     MOM\_{nM} = \sum_{\text{Past} n*21 \text{ days}}  R_{d} - \sum_{\text{Past }  0.1*n*21 \text{ days}}   R_{d}
\end{equation*}
where $R_{d}$ is the daily return.
\subsubsection{Calculate Factor Returns}
For the 12 factors we got, we normalized them cross sections. Then we rank the factor exposures for each sector and take long positions of sectors with the highest two factor exposures, take short position of sectors with the lowest two factor exposures. Then we trade our portfolio under a monthly frequency. Here are the results from 2002 to 2022 February:
\begin{center}
\begin{tabular}{|l|S[table-format=3.1(3)]|S[table-format=3.1(3)]|}
\hline
{Factor}&{Factor Return}&{Sharpe Ratio} \\
\hline
{MOM\_1M} & -0.0297 & -0.09  \\\hline
{MOM\_2M} & -0.0583 & -0.19  \\ \hline
{MOM\_3M} & 0.0082 & 0.02  \\ \hline
{MOM\_4M} & -0.0245 & -0.08  \\ \hline
{MOM\_5M} & -0.0136 & -0.04  \\ \hline
{MOM\_6M} & 0.0195 & 0.06  \\ \hline
{MOM\_7M} & 0.0535 & 0.15  \\ \hline
{MOM\_8M} & 0.0721 & 0.21  \\ \hline
{MOM\_9M} & 0.0413 & 0.11  \\ \hline
{MOM\_10M} & -0.0470 & -0.13  \\ \hline
{MOM\_11M} & 0.0371 & 0.10  \\ \hline
{MOM\_12M} & 0.0327 & 0.01 \\ \hline
\end{tabular}
\end{center}
Since there are several market crashes where the momentum factor led to negative returns, we also take a look at the most recent five years from 2017 to 2022 February:
\begin{center}
\begin{tabular}{|l|S[table-format=3.1(3)]|S[table-format=3.1(3)]|}
\hline
{Factor}&{Factor Return}&{Sharpe Ratio} \\
\hline
{MOM\_1M} & 0.0213 & 0.06  \\\hline
{MOM\_2M} & -0.0494 & -0.14  \\ \hline
{MOM\_3M} & 0.1154 & 0.29  \\ \hline
{MOM\_4M} & -0.1578 & 0.41  \\ \hline
{MOM\_5M} & -0.1788 & 0.47  \\ \hline
{MOM\_6M} & 0.1163 & 0.35  \\ \hline
{MOM\_7M} & 0.2119 & 0.62  \\ \hline
{MOM\_8M} & 0.1869 & 0.56  \\ \hline
{MOM\_9M} & 0.1583 & 0.43  \\ \hline
{MOM\_10M} & 0.0504 & 0.12  \\ \hline
{MOM\_11M} & 0.1547 & 0.40  \\ \hline
{MOM\_12M} & 0.1289 & 0.33 \\ \hline
\end{tabular}
\end{center}
From this table, we can tell that by using the MOM$\_$7M factor, we can reach a maximum annual return rate of 21.19$\%$ and a maximum Sharpe ratio of 0.62. It is also interesting that we find the momentum factor with a short time period, for example, MOM$\_$1M and MOM$\_$2M, have a very small even negative return rate. However, it exactly conforms to the short term reversion effect that the typical momentum factor would exclude. 

\subsection{Short Term Reversion Factor Exploration}
Short term reversion factor follows the simple principle that asset's price will have the trend to stay on an average level. Since we can see from the previous results of the momentum factor that there does exist short term reversion effect, we can try different reversion factors and find out what would be the best short term reversion observation period.

\subsubsection{Factor Construction}
Similarly, we can define several reversion factors with different time periods. And we take the negative number of the past n days cumulative return as the factor exposures.
\begin{equation*}
    REV\_{nD} = - \sum_{\text{Past } n \text{ days}}  R_{d}
\end{equation*}
For the purpose of exploring the optimal time period, we take 5-day time interval and create 12 reversion factors from 5 trade days to 55 trade days.

\subsubsection{Calculate Factor Returns}
By using the same method, we compute the rank of each sector's factor exposure, and long the top two sectors, short the last two sectors on a monthly observation frequency. Between 2002 and 2022 February, the results are:
    \begin{center}
    \begin{tabular}{|l|S[table-format=3.1(3)]|S[table-format=3.1(3)]|}
    \hline
        {Factor} & {Factor Return} & {Sharpe Ratio}  \\ \hline
        {Rev\_5D} & -0.0059 & -0.0597  \\ \hline
        {Rev\_10D} & 0.0294 & 0.3320  \\ \hline
        {Rev\_15D} & 0.0127 & 0.1338  \\ \hline
        {Rev\_20D} & 0.0072 & 0.0715  \\ \hline
        {Rev\_25D} & 0.0690 & 0.7768  \\ \hline
        {Rev\_30D} & 0.0877 & 0.8735  \\ \hline
        {Rev\_35D} & 0.0135 & 0.1288  \\ \hline
        {Rev\_40D} & -0.0314 & -0.3422  \\ \hline
        {Rev\_45D} & -0.0694 & -0.8417  \\ \hline
        {Rev\_50D} & -0.0183 & -0.5919  \\ \hline
        {Rev\_55D} & -0.0517 & -0.1863 \\ \hline
    \end{tabular}
    \end{center}

From this table, we can tell that for the time between 2002 to recent time, the short term reversion effect is optimal for taking the past 30 days cumulative return. It has an optimal annual return rate of 8.77$\%$ on average and leads to a sharp ratio of 0.8735.

\section{Fundamental Analysis}
Fundamental Analysis are always a good aspect to look at for investing. We collected quarterly data for all 11 indices from Bloomberg, including their P/E ratio, EV/EBIT, Profit Margin, etc. Our fundamental analysis would start from discovering features for each of the fundamental ratio,then we are trying to predict the sector performance by constructing using some of the features we found.

\subsection{PE Ratio}
The P/E is one of the most widely used tools to determine a stock’s relative valuation. The purpose of analyzing the ratio is to show whether certain sector is worth to be invested because P/E ratio can reflect the investment risk in this sector. The figure below shows distribution of P/E ratios in different sectors.
\begin{center}
    \includegraphics[scale=0.4]{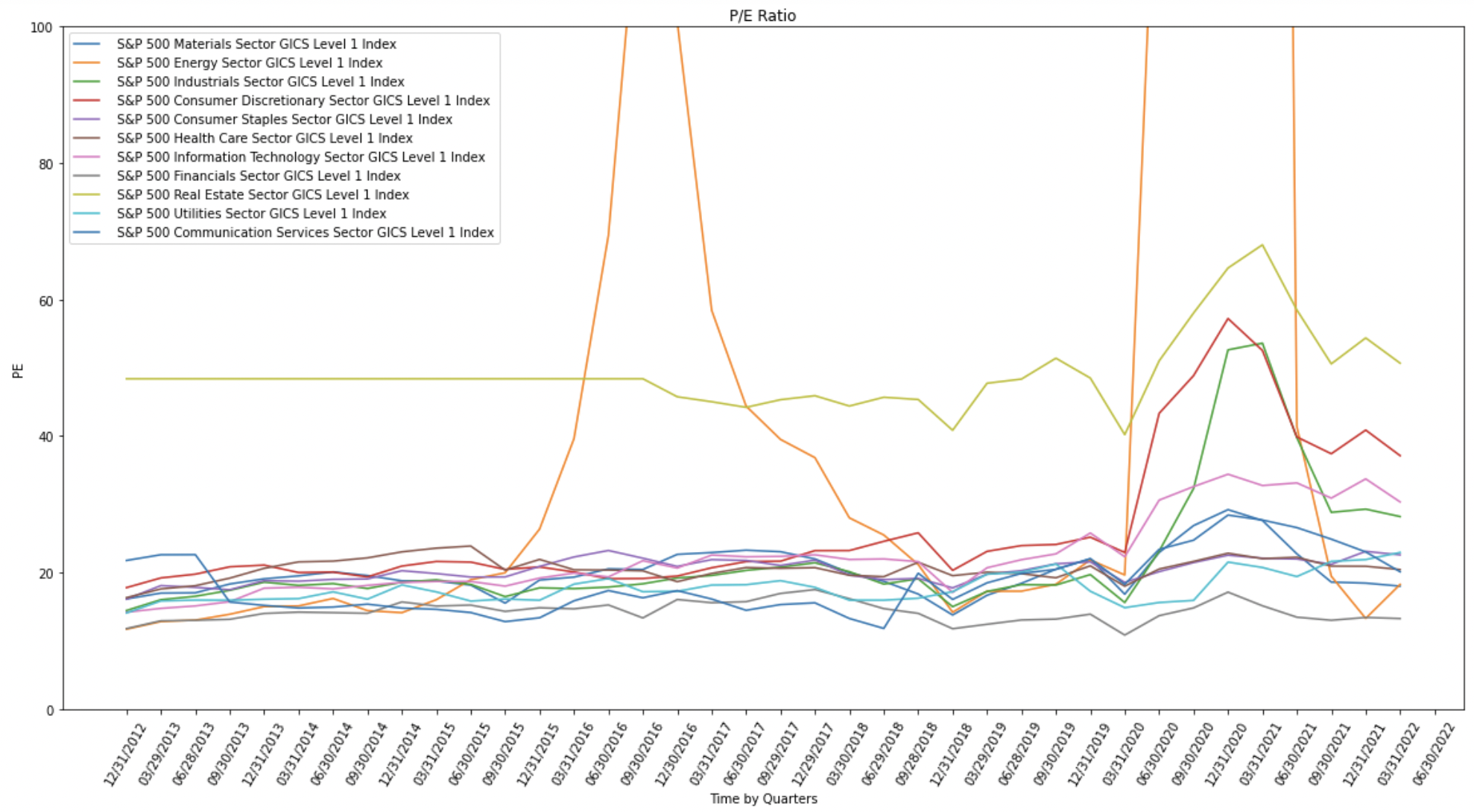}
\end{center}
By comparing cross-sectional data, it is obvious that P/E of Real Estate Sector and Consumer Discretionary Sector are higher than others. The reason is that earning growth in the future is expected to grow fast in the two sectors or these sectors have some special advantages that guarantee long-term profitability with low risk. On the other hand, Financials Sector’s ratio is relatively low compared with other sectors, which may result from its high volatility so investors are reluctant to pay for it.\\
We also notice that the ratio in Energy Sector surged in 2015, which is related to some changes in the sector. The end of the oil age and emergence of alternative energy have reduced the earnings of the original sector. As a result, its relative price becomes higher than before. The change in EV/EBIT and EV/EBITDA is also due to this reason.

\subsection{PB Ratio}
The P/B ratio provides a valuable reality check for investors who are seeking growth at a reasonable price. For those sectors with more assets, their book value and market value are close, so P/B ratio is more useful when we analyze Real Estate sector and Financials sector. The figure below shows distribution of P/B ratio in different sectors.
\begin{center}
    \includegraphics[scale=0.4]{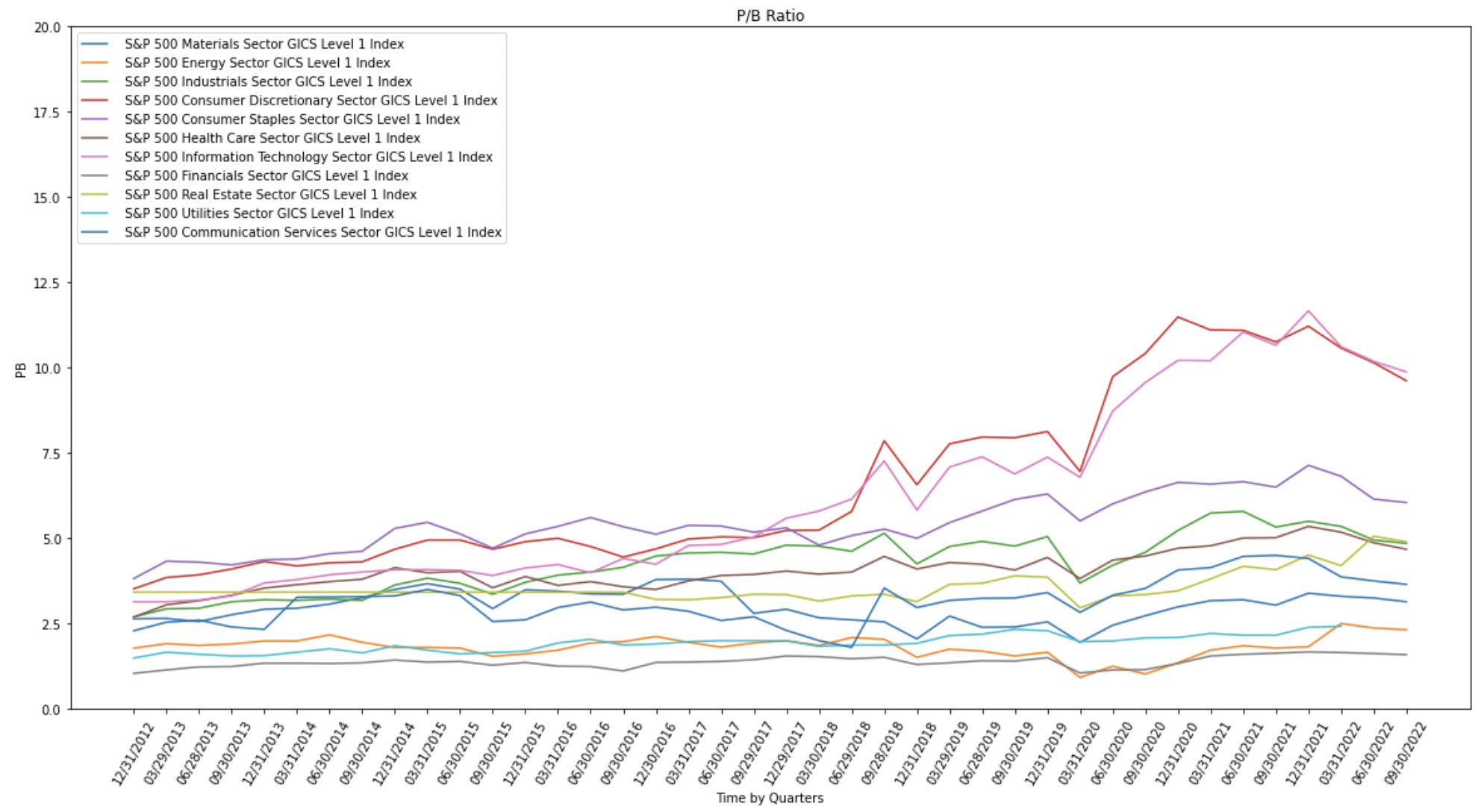}
\end{center}
As the picture shows, Consumer Discretionary sector and IT sector have higher P/B ratio while Financials sector and Energy sector have relatively low ratios. What’s more, Real Estate sector with high P/E ratio has relatively lower P/B ratio.

\subsection{EV/Sales}
EV/Sales can help investors better understand cost relative to unit sales and whether the company is overvalued or undervalued. If EV/Sales is relatively high, the company or sector is less attractive to investors. The figure below shows distribution of EV/Sales in different sectors.
\begin{center}
    \includegraphics[scale=0.4]{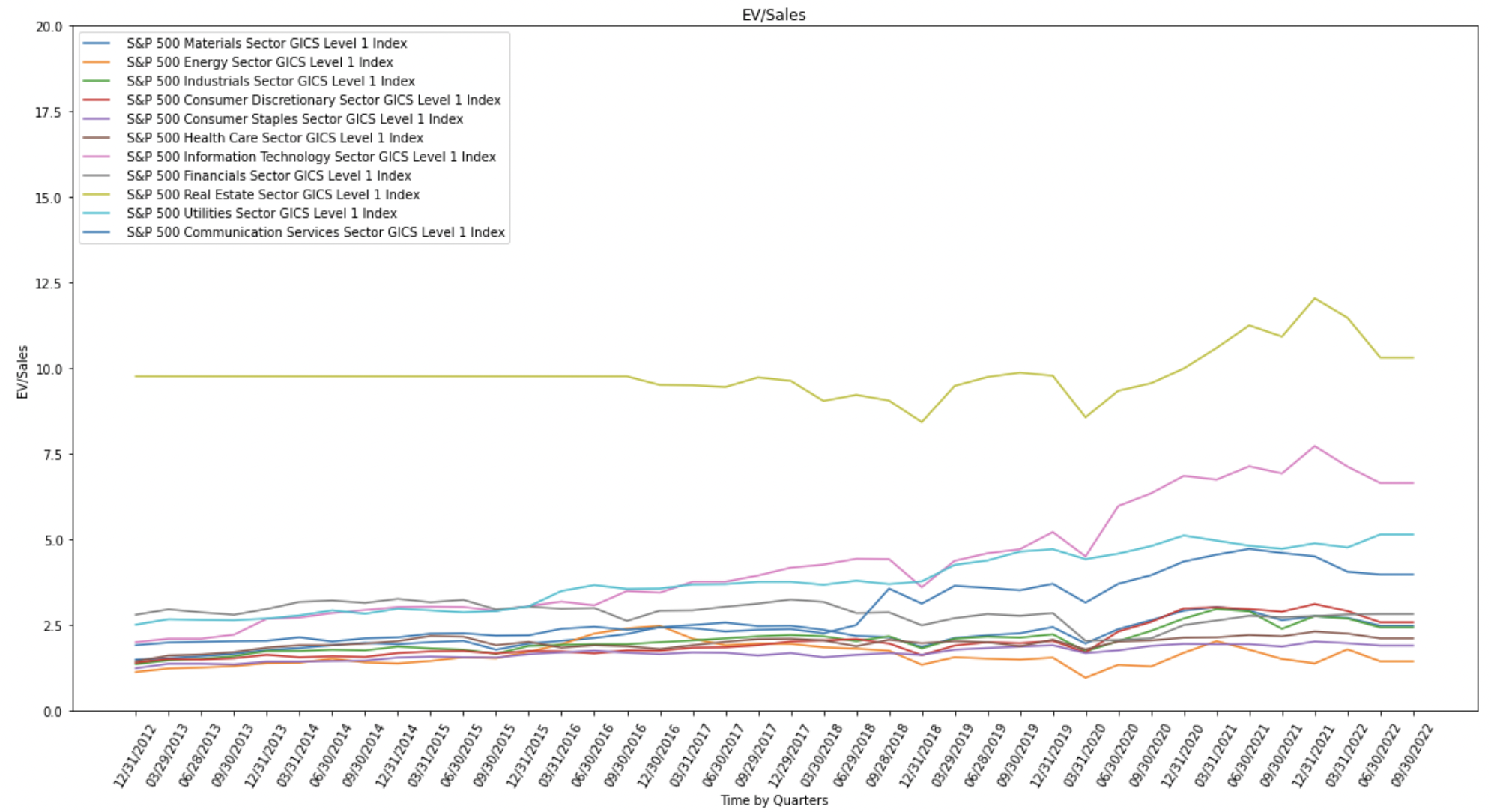}
\end{center}
The result shows that Real Estate sector’s ratio is higher than others’, which means that index in this sector is overvalued. On the other hand, ratio in Energy sector is low, which can attract more investors.

\subsection{EV/EBIT $\&$ EV/EBITDA}
EV/EBIT and EV/EBITDA are independent of the capital structure of the company, whereas multiples like P/E ratio are impacted by financing decisions. Because of this reason, the two are the most commonly relied-upon multiples in relative valuation. However, one obvious distinction is that EV/EBIT considers depreciation and amortization. In some capital-intensive industries which have significant differences in D$\&$A, EV/EBIT may make it a more accurate measure of value. But in our analysis, there is no such difference in the comparison of these two ratios under different sectors.
\begin{center}
    \includegraphics[scale=0.4]{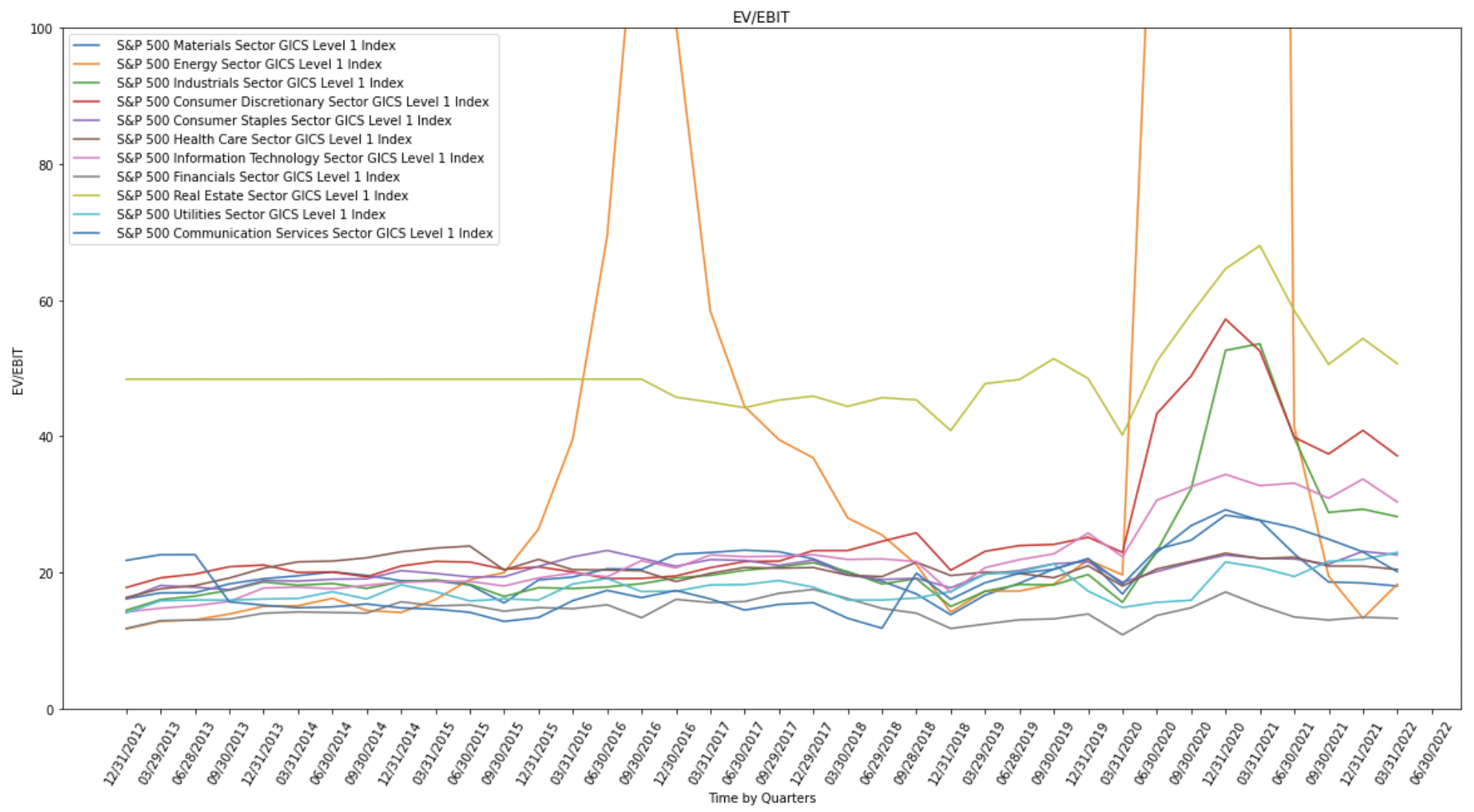}
    \includegraphics[scale=0.4]{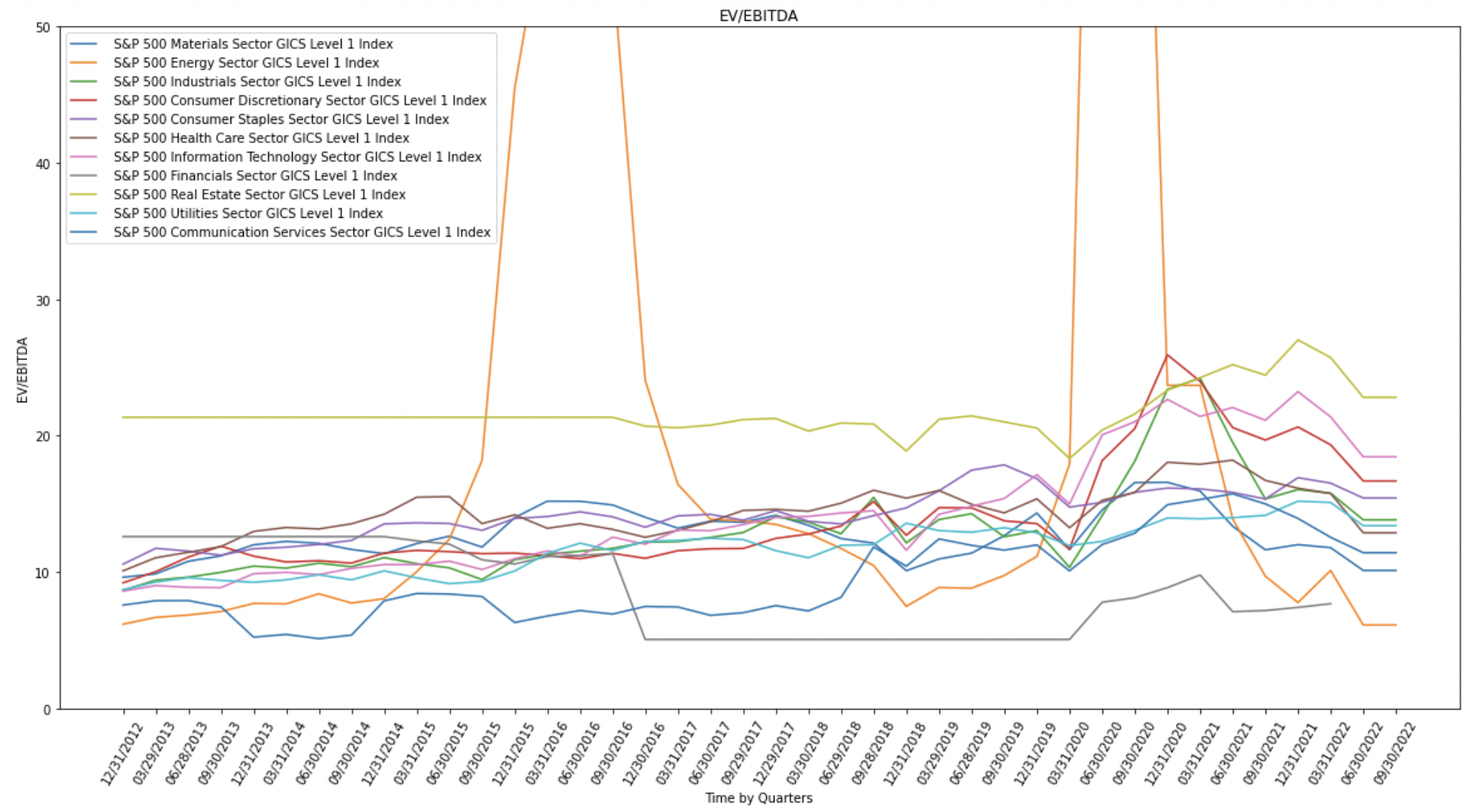}
\end{center}
The results can show that Real Estate sector has a higher ratio. The three ratio EV/Sales, EV/EBIT and EV/EBITDA can give a consensus conclusion that Real Estate sector is overvalued in the market.

\subsection{Dividend Yield}
Dividend Yield is used to measure the amount of cash flow investors are getting back for each dollar. It is essentially the return on investment for a stock without any capital gains. The figure below shows distribution of Dividend yield in different sectors.
\begin{center}
    \includegraphics[scale=0.4]{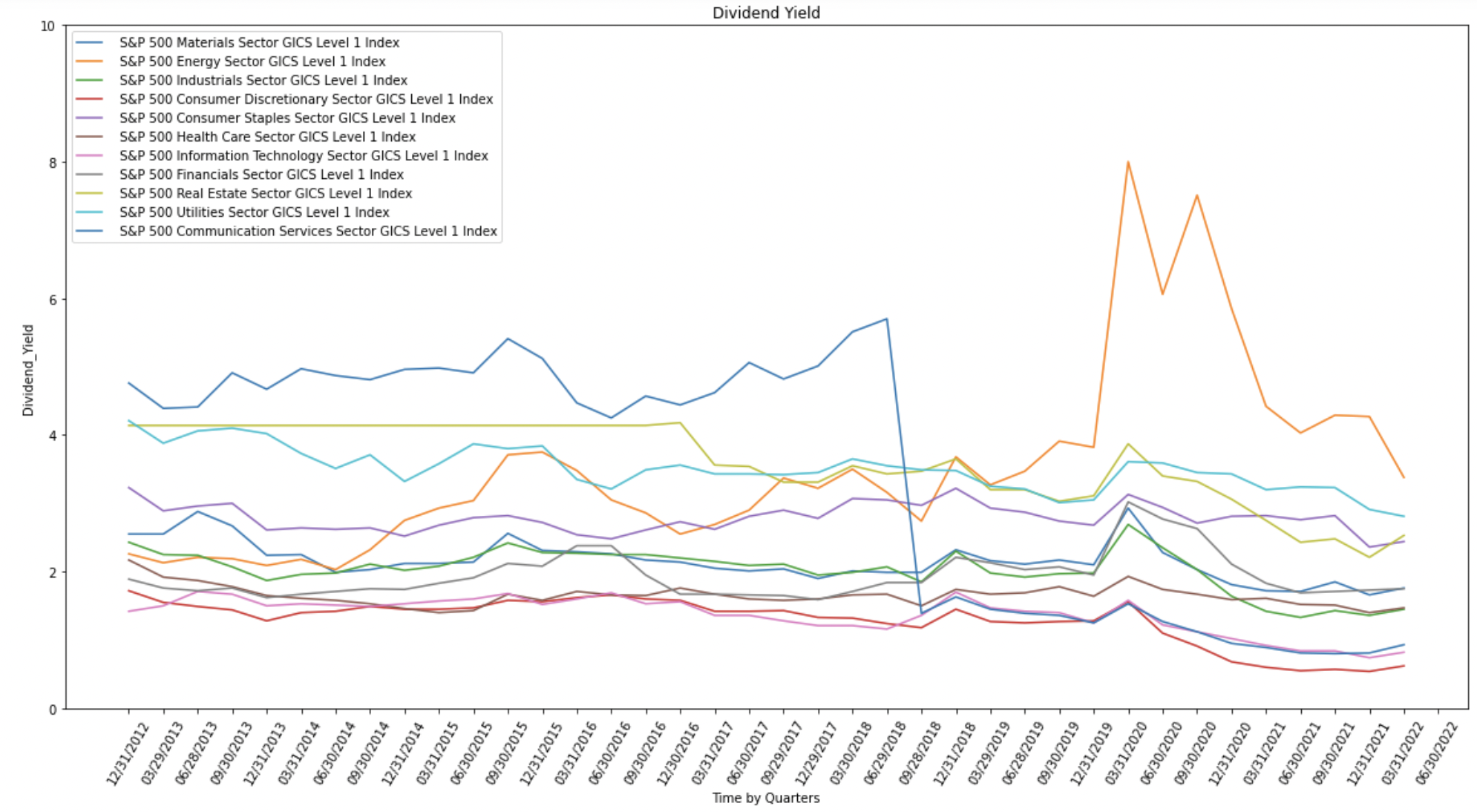}
\end{center}
The ratio in Communication Services sector is higher before 2018 while Energy sector’s ratio is higher after that time. This is because communication services sector took place a reorganization of S$\&$P500 index in 2018. It now includes at least eighteen companies from IT and Consumer Discretionary sectors. Due to this reshuffling, dividend yield of this sector is impacted.

\subsection{Gross Margin}
Gross margin equals net sales less the cost of goods sold (COGS). Net sales are equivalent to the total revenue from sales, and COGS is the direct cost associated with producing goods. By calculating gross margin, we could measure one company's retain revenue after subtracting the production cost. The higher the gross margin, the more capital a company retains, which it can then use to pay other costs or satisfy debt obligations. Generally, companies with good gross margins would have a relatively sustainable competitive advantage. By analyzing gross margin data across sectors, we may observe some sectors that have more stable development in the long run.
\begin{center}
    \includegraphics[scale=0.4]{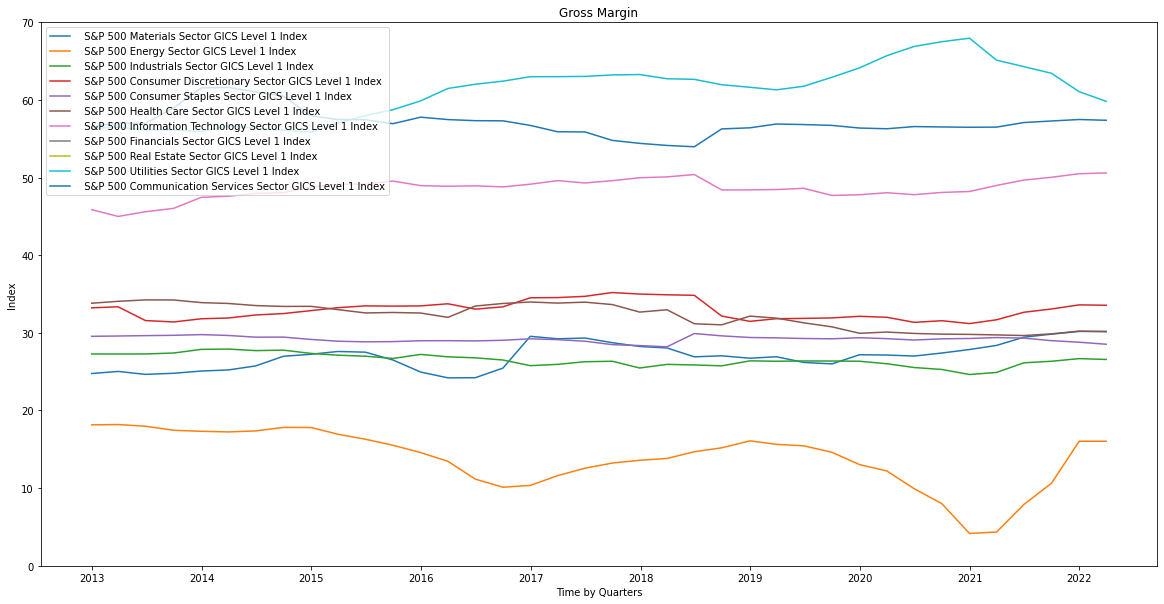}
\end{center}
For our 11 sectors’ gross margin data, the line chart above shows significant differences between the sectors. Overall, each industry index is relatively flat on its own, and have gaps between each others. Utilities, Communication Services and Information Technology(IT) have been among the top spears for last 10 years, occupying the first, second and third positions respectively, all above 40$\%$. On the contrary, the energy sector has been an under-performer for the past decade, ranking at the bottom, with gross margins consistently below 20$\%$. Gross margins in the rest industries are concentrated in the 25$\%$-35$\%$ range and have not fluctuate much.\\
At the same time, by observing the comparison of fluctuations between industries, it is not difficult to see that the gross margin fluctuations of the energy industry and the utilities industry maybe relatively high in the past decade, and their peaks correspond to each other. During 2016, the utilities industry grew significantly, while energy declined comparatively. The trend was even more pronounced in 2020, with utilities reaching its highest level and the energy industry fell to the bottom.\\
Generally, the gross margin feature maybe a significant indicator for Utilities, Communication Services and IT sectors. And our conjecture about the correlation between utilities and energy sectors will need further observation and verification.

\subsection{Operating Margin $\&$ Profit Margin:}
Operating margin equals operating income divided by revenue, it is a profitability ratio measuring revenue after covering operating and non-operating expenses of a business. And profit margin measures the profit ratio after paying for variable costs of production. It is calculated by the formula: 
$$\text{Profit Margin} = \frac{(\text{Total Revenue - Total Expenses )}}{ \text{Total Revenue}}$$
Both operating margin and profit margin are used to gauge the degree of the company's activity makes money. Higher ratios are generally better, illustrating the company is efficient in its operations and is good at turning sales into profits. In our analysis, there is not a very big difference in the comparison of these two ratios under different sectors, which is determined by their definition.
\begin{center}
    \includegraphics[scale=0.4]{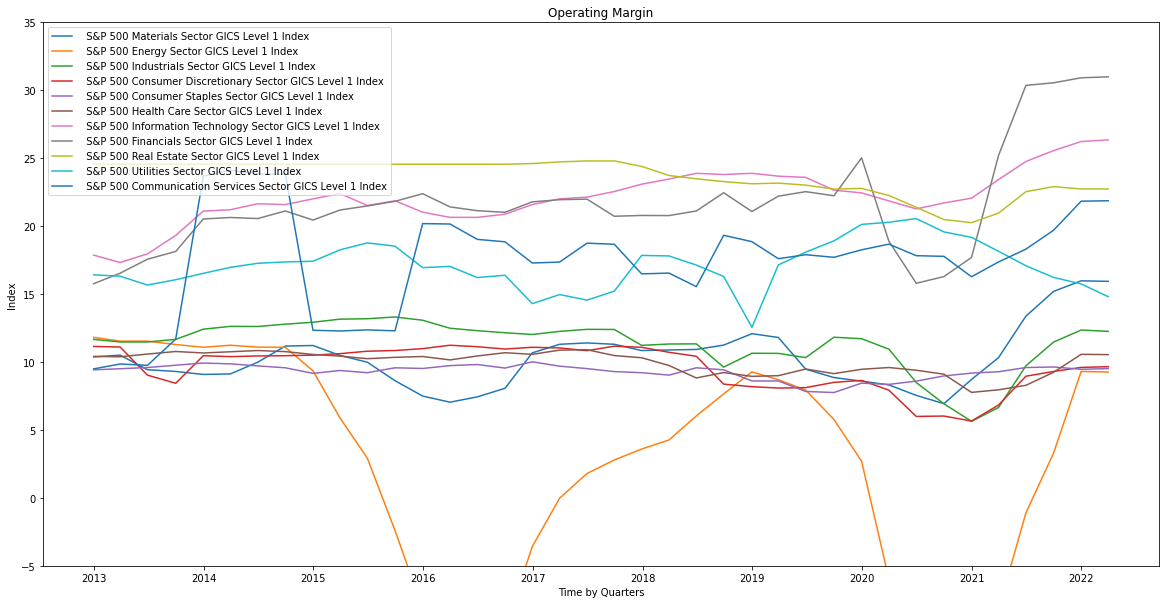}
    \includegraphics[scale=0.4]{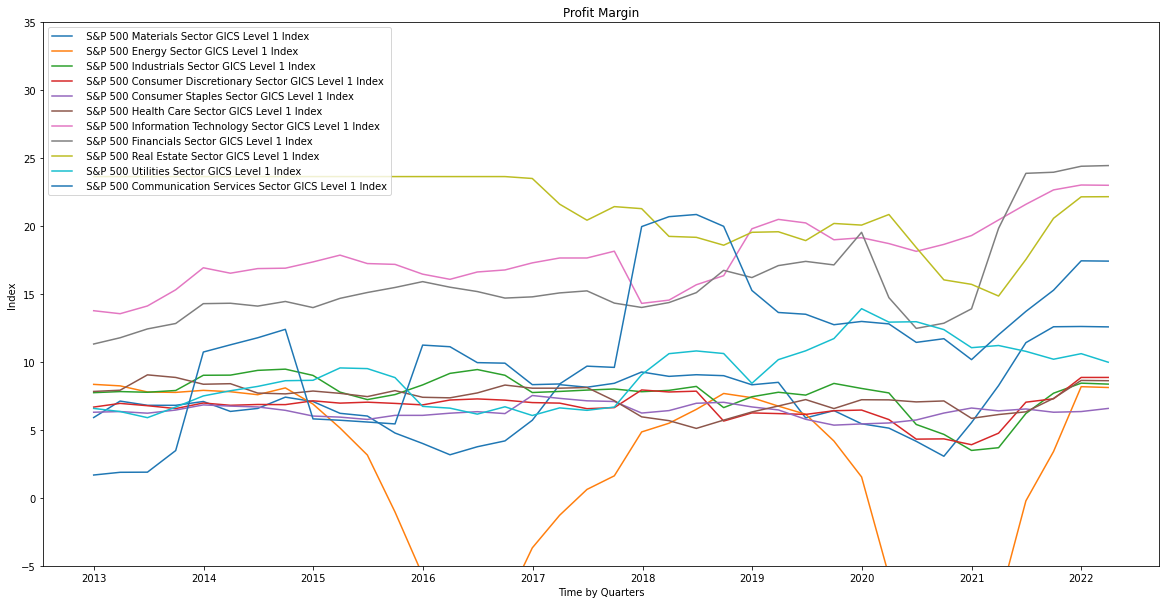}
\end{center}
For these two ratios, Real Estate sector, IT sector and Financial sector have the top three high ratios.And Energy sector has the relatively lowest ratio. Also, both operating margin and profit margin for almost all sectors have similar trends in the last decade curves. This is attributed to the definition difference between the two features, and that's why the operating margin was slightly higher than the profit margin.\\
Another thing that is worth to mentioning is that for Energy Sector, not just operating margin and profit margin, but also the gross margin, it always has the relatively lowest ratios and similar curve fluctuation, with sharp declines in 2016 and 2020. The two time nodes may consistent with some big revolution in the energy industry, which we will analyze later.

\subsection{Return on Asset $\&$ Return on Equity}
Return on equity (ROE) and return on assets (ROA) are two of the most important measures for evaluating how effectively a company’s management team is doing its job of managing the capital entrusted to it. ROE equals to generally net income divided by equity, while Return on Assets (ROA) is net income divided by average assets. So the primary differentiator between ROE and ROA is financial leverage or debt. ROE measures profitability and ROA is an efficiency measure of how well a company is using its assets.
Investors may prefer to observe ROE, since equity represents the owner's interest in the business. Compared to other sources of fund, equity capital tends to be the most expensive source of funding and carries the largest risk premium of all financing options. Therefore, in our analysis, ROE may be a better feature that it could reflect the trend of market investment.
\begin{center}
    \includegraphics[scale=0.4]{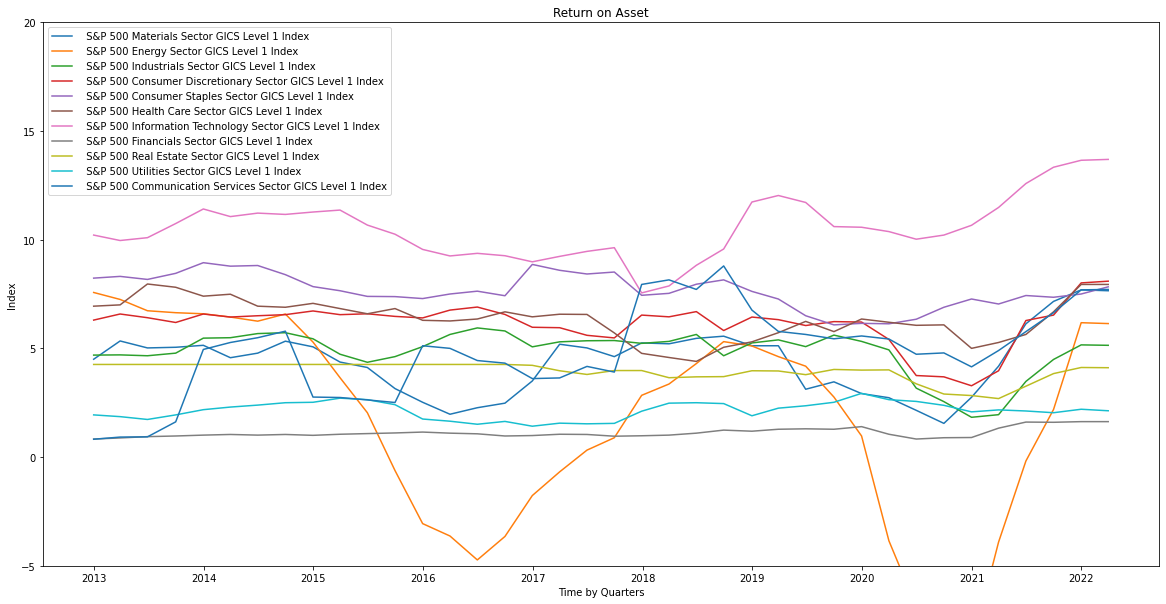}
\end{center}
As shown in the picture, IT sector has the highest ROA, the Consumer Staples sector and Consumer Discretionary sector also have a relatively higher ratio. In contrast, Financial sector has a lower ROA. 
The past ten years, or even twenty years, has been an era of rapid development of information technology. And compared with traditional industry and commerce, information technology is more flexible in the time and form of investment assets, that's the reason why IT will have the highest ratio. Also for the the Consumer Staples sector and Consumer Discretionary sector，they are all industries with fast innovation and short production cycle. Generally, these three will have constantly higher ratio for the long run. Therefore, for these three industries, if the ROA indicator fluctuates significantly, it may have an impact on the investment trend.
\begin{center}
    \includegraphics[scale=0.40]{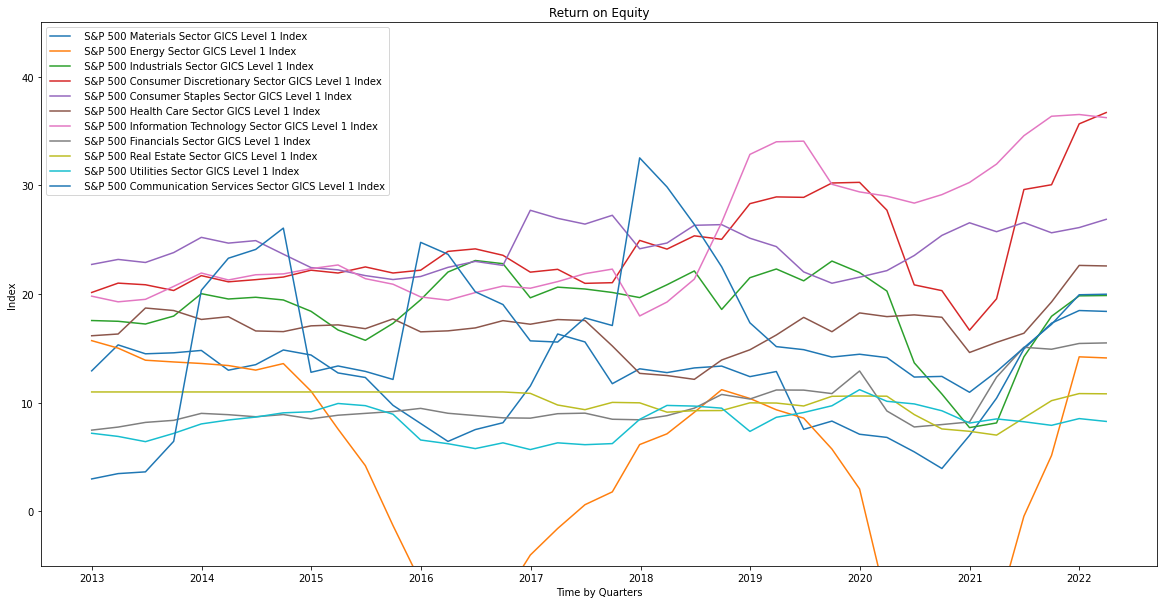}
\end{center}
For ROE ratio, similarly, IT, Consumer Staples stay high, and Consumer Discretionary sectors is also at a slightly higher level, except that the IT sector lost its prominence in ROA ratio. 
By comparing cross-sectional data, the Consumer Discretionary Sector and Industrials Sector have similar patterns in the last decade for both ROA and ROE ratios. They both have a low peak in 2020. It is conceivable that this is affected by the general environment of the epidemic.
And as we mentioned before, the ROE and ROA curves of the energy sector still have a similar pattern, falling sharply in 2016 and 2020. In 2016, it was affected by changes in energy policy since 2015, reducing oil production while encouraging the development of clean and new energy. For 2020, we attribute this decline to the outbreak of the COVID-19 pandemic.
\section{Prediction by Fundamental Factors}
\subsection{Factor and Future Return}
Having these fundamental data, next step is to find out what quantitative relationships they have to futures sector returns. For fundamental factors, they are usually exposed in the company report with annual, semi-annual, or quarterly frequency. Our fundamental factors for each sector are reported quarterly, leading to a problem that the sample size for each individual sector is very small. To have a better performance of the prediction model, we need to combine all the sectors together and make a uniformed and comparable large sample. We neutralized each factor cross-sectional for the factor to have a mean of 0 and standard deviation of 1. If $X_{i,t}$ denotes one specific factor exposure for $i$-th sector at time $t$, in this case would be at $t$-th quarter, then for each individual $t$ we have the neutralized exposure to be:
$$X_{i,t}\text{ -Neutral}=\frac{X_{i,t}-\text{Mean}(X_{i,t}\text{  ,}i \text{ from } 1 \text{ to } 11)}{\text{Standard Deviation}(X_{i,t}\text{  ,}i \text{ from } 1 \text{ to } 11)}$$
Then we used the next quarter's cross-sectional normalized return as the corresponding return. First, we want to have a general view of the relations. The scatter plots between neutralized factors and future returns are as following:
\begin{center}
\includegraphics[scale=0.5]{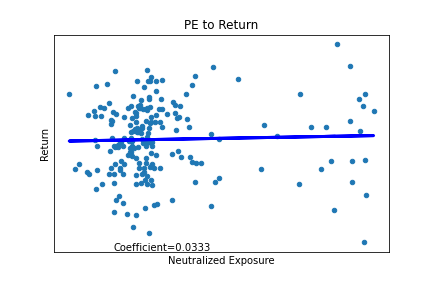}
\includegraphics[scale=0.5]{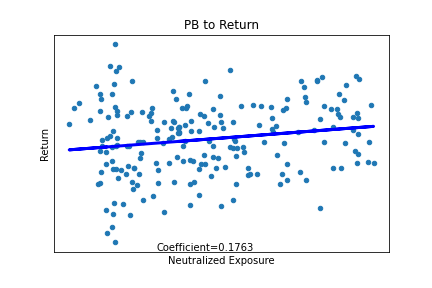}\\
\includegraphics[scale=0.5]{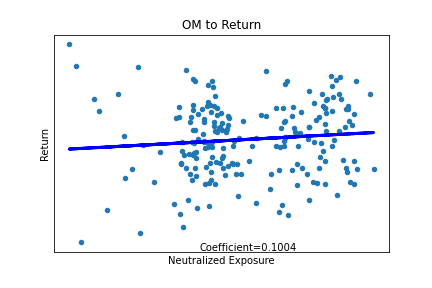}
\includegraphics[scale=0.5]{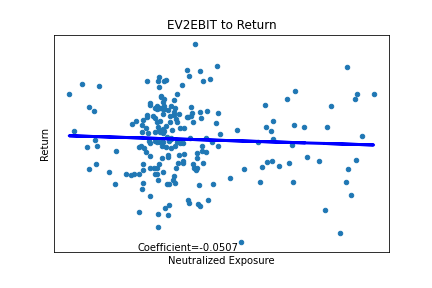}\\
\includegraphics[scale=0.5]{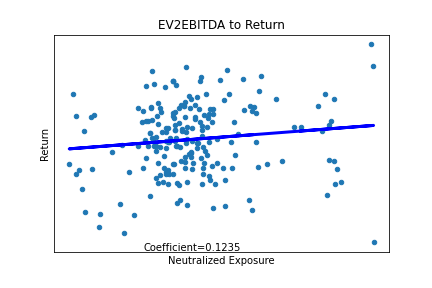}
\includegraphics[scale=0.5]{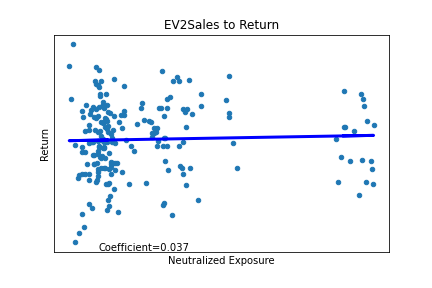}\\
\includegraphics[scale=0.5]{rOM.png}
\includegraphics[scale=0.5]{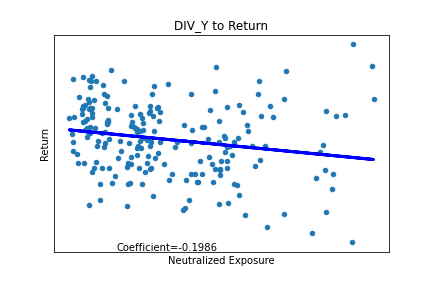}\\
\includegraphics[scale=0.5]{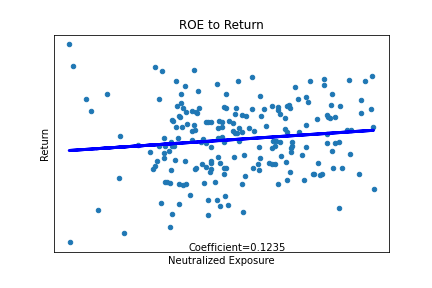}
\includegraphics[scale=0.5]{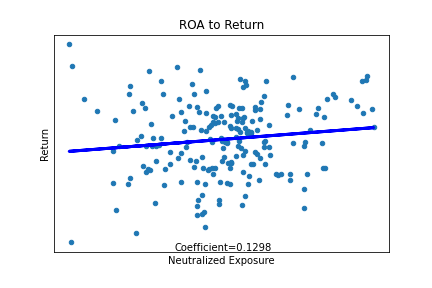}\\
\end{center}
From the scatter plots, the relations between all factors and their future returns cannot be well interpreted by simple linear models. However, it is very common in the financial field that the sample will have a very low signal-noise ratio.
\subsection{Model Construction and Training}
As we observed before, the relations between each factor and its future return cannot be interpreted very well by linear models. Also, we have no idea what model would exactly best fit the data. Therefore, converting prediction of future returns to a classification problem and fitting the training sample with a neural network model which has comparably good performance with non-linear relations would be a great start point. 
\begin{center}
    \begin{neuralnetwork}[height=10]
            \newcommand{\x}[2]{$x_#2$}
            \newcommand{\y}[2]{$\hat{y}_#2$}
            \newcommand{\hfirst}[2]{\small $h^{(1)}_#2$}
            \newcommand{\hsecond}[2]{\small $h^{(2)}_#2$}
            \inputlayer[count=10, bias=false, title=Input\\layer, text=\x]
            \hiddenlayer[count=5, bias=false, title=Hidden\\layer 1, text=\hfirst] \linklayers
            \hiddenlayer[count=5, bias=false, title=Hidden\\layer 2, text=\hsecond] \linklayers
            \outputlayer[count=2, title=Output\\layer, text=\y] \linklayers
    \end{neuralnetwork}
\end{center}
Neural network takes a vector as the input, and goes to each of the neuron in the first hidden layer and gains new activation vectors which act as the input for next hidden layer. After the last hidden layer, neural network model would pass out the probability for each of the prediction class and we choose the one with the highest probability as the prediction. This process is called front propagation. After comparing the prediction to the actual results, we adjust the weights of the nodes by using back propagation for each training pair in the training samples. Also, we use the rectified linear unit function as the activation function for hidden layers and sigmoid function as the activation function for final output. Since we only have a sample of size 200, choosing quasi-Newton methods as the solver has better performance for small sample training.
Then we need to construct the training, validation, and test sets. Since the fundamental factors are already neutralized (normalized) within each sector, we divide the sample data to 60$\%$, 20$\%$, 20$\%$ by convention. Without shuffling, we will have the historical data divided where test set contains the most recent data. For the corresponding output value, we assign 1 to samples with positive future return and 0 with negative returns.
The complexity of neural network directly related to the number and sizes of hidden layers. For the purpose of avoiding overfitting or under-fitting, we need to find proper hyper parameters for neural network model. We start from a simple model with two layers. Let $N$ denote the number of nodes in each hidden layer, alpha is the hyper parameter for L2 regularization penalty function. With larger $N$, the model is more complex. If alpha increases, the penalty for large weights increases, which makes the model tend to be more simple. Considering our sample size is small, intuitively we need to focus more on the overfitting problem. For a range of alpha and $N$, we train the model using the training set data, and get the score for prediction on validation set. The score represents the probability of making a right prediction. Here are the results:
    \begin{center}
    \begin{tabular}{|c|c|c|c|c|c|c|c|c|c|c|}
   \hline
        N$\backslash$alpha & 3 & 2.5 & 2 & 1.5 & 1 & 0.5 & 0.25 & 0.1 & 0.01 & 0.001  \\ \hline
        5 & 0.49 & 0.45 & 0.49 & 0.56 & 0.62 & 0.58 & 0.62 & 0.58 & 0.56 & 0.56  \\ \hline
        6 & 0.51 & 0.55 & 0.45 & 0.49 & 0.51 & 0.49 & 0.47 & 0.56 & 0.56 & 0.56  \\ \hline
        7 & 0.55 & 0.53 & 0.56 & 0.51 & 0.47 & 0.51 & 0.58 & 0.49 & 0.53 & 0.57  \\ \hline
        8 & 0.51 & 0.47 & 0.55 & 0.51 & 0.51 & 0.64 & 0.6 & 0.6 & 0.59 & 0.6  \\ \hline
        9 & 0.53 & 0.49 & 0.47 & 0.53 & 0.56 & 0.56 & 0.55 & 0.6 & 0.53 & 0.6  \\ \hline
        10 & 0.55 & 0.53 & 0.53 & 0.53 & 0.51 & 0.51 & 0.53 & 0.6 & 0.6 & 0.56  \\ \hline
        11 & 0.53 & 0.58 & 0.51 & 0.51 & 0.47 & 0.65 & 0.58 & 0.55 & 0.56 & 0.51  \\ \hline
        12 & 0.55 & 0.55 & 0.56 & 0.51 & 0.49 & 0.51 & 0.54 & 0.53 & 0.53 & 0.53  \\ \hline
        13 & 0.53 & 0.47 & 0.55 & 0.49 & 0.56 & 0.6 & 0.55 & 0.58 & 0.6 & 0.56  \\ \hline
        14 & 0.49 & 0.51 & 0.56 & 0.53 & 0.55 & 0.53 & 0.49 & 0.51 & 0.6 & 0.53  \\ \hline
        15 & 0.47 & 0.51 & 0.51 & 0.51 & 0.45 & 0.49 & 0.56 & 0.6 & 0.56 & 0.55 \\ \hline
    \end{tabular}
\end{center}
To better understand how the hyperparameters influence model performance, we visualize the data by using $N$ and alpha as the bottom coordinates, and use the corresponding probability as the height.
\begin{center}
\includegraphics[scale=0.4]{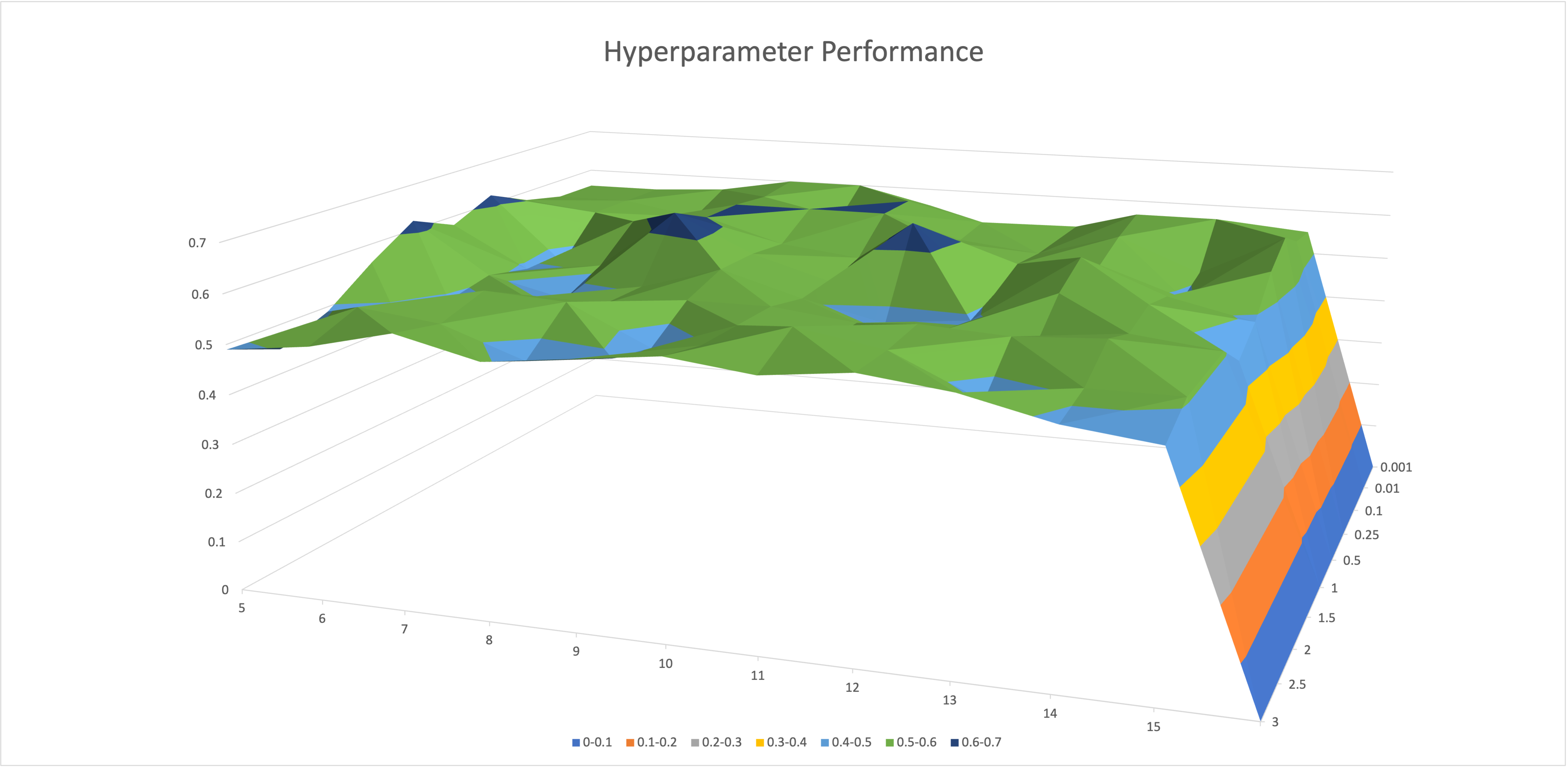}
\end{center}
From the figure, we can tell that the model have several local optimal pairs. And the scores at the optimal points with relative large N values are also combined with small alpha values. For example, the combination of 14 nodes and alpha equals 0.01 has a local optimal score of 0.6. Since we are training with a small sample, using such a complex model with a high score is highly likely overfitting. Therefore, we start from the simple model by looking at models with 5 nodes model and check how the score varies with alpha. For model with 5 nodes, we see there are two local peaks with alpha equal to 1 and 0.25, then we pick the middle value 0.5 as the value of alpha considering the trade-off between variance and bias. 

\subsection{Model Performance}
Constructed and trained the model, next we would test the model by feeding a new data set to the model. On the test set, the score of the model is 0.64, which means the model predicts 64$\%$ of the results correctly. More detailed results are showed in the following table:

\begin{table}[!ht]
    \centering
    \begin{tabular}{|l|c|c|}
    \hline
        ~ & Positive Return& Negative Return\\ \hline
        Actual Sample & 27 & 28 \\ \hline
        Prediction & 37 & 18 \\ \hline
        Correct Prediction & 22 & 13 \\ \hline
        Winning Rate & 0.59 & 0.72 \\ \hline
    \end{tabular}
\end{table}
On the test set, we have a 0.59 winning rates on the positive predictions and 0.72 on the negative predictions, which gives an overall winning rate of 0.64.

\subsection{Trade with Model Prediction}
By using the predictions from the trained model, we used the data from validation set to get trade signals. Instead of having signals of 1 or 0 as the model's output, we choose the probability of the prediction output being 1, which is given by the activation sigmoid function. Then we will have a time series of the probability for each sector, and rank the probability from highest to lowest where the highest probability will have a rank 1. For each cross-sectional ranking, we equally-weighted long sectors with rank 1 to 3 and short sectors with rank 9 to 11 to construct a dollar-neutral portfolio. On the test set, which is from September in 2020 to September in 2021, we have a Sharpe ratio of 2.21. The cumulative return plot is following:
\begin{center}
\includegraphics[scale=0.4]{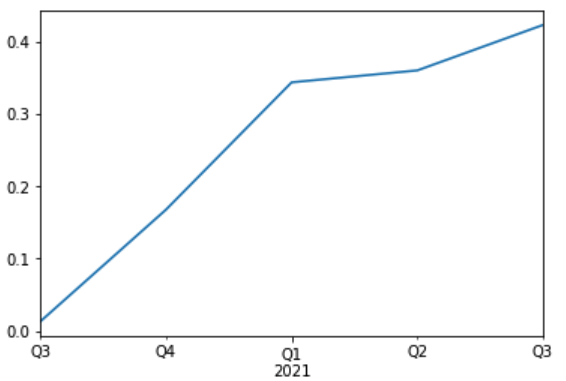}
\end{center}
\subsection{Other Considerations}
There are still issues that need to be considered carefully in the future. First is the factor neutralization. In previous model, we neutralized the factor exposure cross-sectionally, where the exposures reflect the relative level of factor exposure for one sector compared to other sectors at a given time. However, different sectors may have inner trends of higher exposures than others for some factors, especially for fundamental factor. What's more, we only have quarterly fundamental data available from 2017 and it is hard to implement time series normalization for each sector. Therefore, how to modify the factor exposures to make them comparable is a difficult problem. Secondly, as the sample size is small, the model might not be applicable on a wider range of time since we only trained and tested on the most recent five years. One possible way to improve this model is to use daily factors such as volume, close price as input, and convert fundamental factors to daily frequency by the corresponding quarter. Then we would have a sample size of approximately 1250 for each sector and over 13000 samples for training. However, the model might depends more on the daily factors rather than fundamental factors since their exposures would be the same value for each quarter.
\section{References}
1. \textit{Returns to Buying Winners and Selling Losers: Implications for Stock Market Efficiency} Narasimhan Jegadeesh; Sheridan Titman
The Journal of Finance, Vol. 48, No. 1. (Mar., 1993), pp. 65-91.\\
2.The Global Industry Classification Standard, MSCI (1999)
\end{document}